\documentclass[12pt]{article}
\begin{document}
\def\parno{\par \noindent}
\def\lr #1{\mathrel{#1\kern-.75em\raise1.75ex\hbox{$\leftrightarrow$}}}
\def\bi{\begin{itemize}}
\def\ei{\end{itemize}}
\def\benu{\begin{enumerate}}
\def\eenu{\end{enumerate}}
\def\be{\begin{equation}}
\def\ee{\end{equation}}
\def\ba{\begin{eqnarray}}
\def\ea{\end{eqnarray}}
\def\nn{\nonumber}            
\def\R{{\hbox{{\rm I}\kern-.2em\hbox{\rm R}}}}   
\def\H{{\hbox{{\rm I}\kern-.2em\hbox{\rm H}}}}   
\def\N{{\hbox{{\rm I}\kern-.2em\hbox{\rm N}}}}   
\def\C{{\ \hbox{{\rm I}\kern-.6em\hbox{\bf C}}}} 
\def\Z{{\hbox{{\rm Z}\kern-.4em\hbox{\rm Z}}}}   
\def\downarcfill{${\mathsurround=0pt }
\braceld\leaders\vrule\hfill\bracerd $}
\def\arc#1{\mathop{\vbox{\ialign{##\crcr\noalign{\kern3pt}
\downarcfill\crcr\noalign{\kern3pt\nointerlineskip}
$\hfil\displaystyle{#1}\hfil$\crcr}}}\limits}

\def\mettresous#1\sous#2{\mathrel{\mathop{\kern0pt #2}\limits_{#1}}}
\def\sqr#1#2{{\vcenter{\vbox{\hrule height.#2pt
          \hbox{\vrule width.#2pt height#1pt \kern#1pt
           \vrule width.#2pt}
           \hrule height.#2pt}}}}
\def\dalamb{\mathchoice\sqr68\sqr68\sqr{4.2}6\sqr{3}6}
\def\ket#1{|#1\rangle}      
\def\bra#1{\langle #1|}     
\def\kvac {|0\rangle}                   
\def\bvac{\langle 0|}       
\def\braket#1#2{\mathrel{\langle #1|#2\rangle}}   
\def\elematrice#1#2#3{\langle #1|#2|#3 \rangle}   
\def\inoutexpect#1{\elematrice{0,\mbox{out}}{#1}{0,\mbox{in}}}
\def\inout{\langle 0,\mbox{out}|0,\mbox{in}\rangle}

\newlength{\fleche}
\newcounter{FLECHE}
\newcommand{\vecteur}[1]
{
\settowidth{\fleche}{$#1$}
\setcounter{FLECHE}{\fleche}
\setlength{\unitlength}{1sp}
\stackrel
{\begin{picture}(\value{FLECHE},0)
\put(0,0){\vector(1,0){\value{FLECHE}}}
\end{picture}
}
{#1}
\setlength{\unitlength}{1pt} }

\newcounter{DEMIFLECHE}
\newcommand{\lrvect}[1]
{\settowidth{\fleche}{$#1$}
\setcounter{FLECHE}{\fleche}
\newcount\demifleche
\demifleche=\value{FLECHE}
\divide\demifleche by 2
\setcounter{DEMIFLECHE}{\demifleche}
\setlength{\unitlength}{1sp}
\stackrel
{\begin{picture}(\value{FLECHE},0)
\put(\value{DEMIFLECHE},0){\vector(1,0){\value{DEMIFLECHE}}}
\put(\value{DEMIFLECHE},0){\vector(-1,0){\value{DEMIFLECHE}}}
\end{picture} }
{#1} \setlength{\unitlength}{1pt} }

\def\lrpartial{\mathrel{\partial\kern-.75em\raise1.75ex\hbox{$\leftrightarro
w$}}}
\def\lrD{\mathrel{{\cal D}\kern-.75em\raise1.75ex\hbox{$\leftrightarrow$}}}
\def\mettresous#1\sous#2{\mathrel{\mathop{\kern0pt #2}\limits_{#1}}}
\def\zentier{\ \hbox{{\rm Z}\kern-.4em\hbox{\rm Z}}}
\def\reel{\ \hbox{{\rm R}\kern-1em\hbox{{\rm I}}}}
\def\bfGamma{\ \hbox{{$\Gamma$}\kern-.5em\hbox{{\rm I}}}\,}
\def\Iden{\ \hbox{{1}\kern-.25em\hbox{{\rm I}}}\,}
\def\souligne#1{$\underline{\smash{\hbox{#1}}}$}
\def\1op{\ \hbox{{\rm 1}\kern -0.23em\hbox{{\rm I}}}}
\def\ov{\overrightarrow}
\def\spind{\vcenter{\hsize=1em \baselineskip 0pt \parindent 0pt
$\to$ \\ $O$ \\ $\to$}}
\def\spindo{\vcenter{\hsize=1em \baselineskip 0pt \parindent 0pt
$\to$ \\ $\omega$ \\ $\to$}}
\def\spindi{\vcenter{\hsize=1em \baselineskip 0pt \parindent 0pt
$\to$ \\ $I$ \\ $\to$}}
\newcommand{\figdef}[4]
{\begin{figure}
\vglue10pt
\vspace{#1}
\special{illustration #2 scaled 800}
\caption{{\sl #3}}
\vspace{3mm}
\label{#4}
\end{figure}
}
\catcode`\@=11
\@addtoreset{equation}{section}
\def\theequation{\arabic{section}.\arabic{equation}}
\def\appendix{\renewcommand{\thesection}{\Alph{section}}\setcounter{section}{0}
              \renewcommand{\theequation}
            {\mbox{\Alph{section}.\arabic{equation}}}\setcounter{equation}{0}}
\renewcommand{\thesection}{\Roman{section}}
\renewcommand{\thesubsection}{\thesection-\Alph{subsection}}

\vskip 1. cm
\begin{center} {\LARGE\bf   Thermodynamic implications of some unusual quantum theories.}\\
\vskip 1. cm M. Lubo\footnote{E-mail lubo@sun1. umh. ac.
be}$\mbox{}^{\mbox{\footnotesize }}$,
\\ {\em M\'ecanique et Gravitation, Universit\'e de Mons-Hainaut},
\\ {\em 6, avenue du Champ de Mars, B-7000 Mons, Belgium}
 \\
\vskip 0.5 cm
\vskip 1 cm
\end{center}
\begin{abstract}
\par
Various deformations of the position-momentum algebras operators
have been proposed. Their implications for single systems like the
hydrogen atom or the harmonic oscillator have been addressed. In
this paper we investigate the consequences of some of these
algebras for macroscopic systems. The key point of our analysis
lies in the fact that the modification of the Heisenberg
uncertainty relations present in these theories changes the volume
of the elementary cell in the hamiltonian phase space and so the
measure needed to compute  partition functions.
\par The thermodynamics of a non interacting gas are studied for two
members of the Kempf-Mangano-Mann (K.M.M.) deformations. It is shown
that the theory which exhibits a minimal  uncertainty  in length predicts a new
behavior at high temperature while the one with a minimal uncertainty in
 momentum displays  unusual features for huge volumes. In the second  model negative
 pressures are obtained and mixing two different gases
  does not  necessarily increase  the entropy  .
 This suggests   a possible violation of the second law of
thermodynamics.
  Potential consequences of these models in the  evolution of the early universe
  are briefly discussed.
\par Constructing the Einstein model of a solid for the $q$ deformed
oscillator, we find that the subset of eigenstates whose energies
are bounded from above  leads to a divergent partition function.

\end{abstract}
\vfill
\newpage
\section{Introduction}
\par Modern physics is an edifice in which every stone is tightly linked
to the others. A slight modification in one area may produce important
changes in  different fields.
 Quantum mechanics starts with the commutation relations. Once they have been fixed,
 using  a representation of the Heisenberg algebra we can in principle solve the
 Schrodinger equation whose solutions potentially contain all the physics of
  the ( non relativistic) system studied. The correspondence principle which states the link
between the commutator of two variables of the phase space and
their classical Poisson bracket is one of the basic axioms of
quantum mechanics. It is not deduced from another assumption and
cannot be judged in an isolated manner : only the whole theory can
be confronted to experience through its predictions.
\par Some authors have investigated the consequences of the alterations of
these commutation relations on the observable of some physical
systems \cite{LABEL1} - \cite{LABEL11}.
 In particular, some deformations of the canonical variable commutators
studied by Kempf-Mangano-Mann(K.M.M)  induce a minimal uncertainty
in the position (or the momentum) in a very simple way, providing
a toy model with manifest non locality. The implications of some
of these quantum structures have been studied for the harmonic
oscillator \cite{LABEL1,LABEL3} and the hydrogen atom
\cite{LABEL12}. The transplanckian problem occurring in the usual
description of the
 Hawking mechanism of black hole evaporation has also been addressed in this framework,
 for the Schwarzchild
and the Banados-Teitelboim-Zanelli(B.T.Z) solutions \cite{LABEL5,LABEL6}.

\par The purpose of this paper is the investigation of the modifications
these models induce, not in the characteristics of a single
particle, but in the behavior of a macroscopic system. The modern
presentation of thermodynamics basically relies on statistical
physics. According to the system under study (isolated, closed or
open), one uses an ensemble (microcanonical , canonical or grand
canonical) and the corresponding potential (entropy, free energy,
grand potential) to derive thermodynamic quantities (pressure,
specific heat,  chemical potential, etc...).
\par What has this to do with the commutation relations? The thermodynamic
potentials are related to the derivatives of the partition
function and the partition function itself is an average (of a
quantity which depends on the ensemble used) on the phase space.
To define a measure on the phase space, one needs to know the
extension of the fundamentals cells. In the wrongly called
``classical'' statistical Physics, the Heisenberg uncertainty
relations are used to show that this volume is  the cube of the
Planck constant and the indiscernability of particles justifies
the Gibbs factor.
 The difference with quantum statistical physics lies in the
fact that in the last case the sum giving the partition function  has to be made on a
discrete set of
states. For example, to explain the specific heat of a solid, the
Einstein model treats each atom as a quantized oscillator.
\par  If one modifies the
commutators, one changes the energy spectrum and the Heisenberg
uncertainty relations. The measure on the phase space is no more
the same and this results in new partition functions and
consequently different thermodynamic behaviors .

\par The paper is organized as follows. In the second section we give
a brief remind of the basics of statistical physics (to fix the notations)and their application
in the derivation of the thermodynamic properties of some simple systems.
The third section begins with a short overview of K.M.M. deformations
of the momentum-position commutator. The non interacting gas  is studied in
two special cases: when a minimal uncertainty in position or in momentum
is present , in the non relativistic
and the ultra relativistic regimes. The Einstein model describing
a crystal is also investigated. The last section introduces the harmonic
oscillator of the $q$ deformed algebra and uses it to work out the
Einstein model of a solid in this particular theory.

\section{Basics of statistical physics.}
  \par Let us consider, in the  usual theory, a system which is in contact with a
 large heat reservoir and doesn't exchange particles with the surroundings :
it has to be studied in the canonical ensemble \cite{LABEL13,
LABEL14}. Its equilibrium state will be described by a fixed
temperature and a fixed particle number while its energy will
fluctuate around a mean value. Strictly speaking, for such a
system , the particle number $ N $ is fixed once and for all. But,
one knows that when phase transitions are not present, the
descriptions given by the canonical and the grand canonical
ensembles are very close. This will be used to compute the
chemical potential in the canonical ensemble which is more
tractable.
   Taking for example  an assembly of non interacting non relativistic
particles in a cubic box of length $L$, one  solves the
Schr\"Schrodinger equation \be \label{2.1}
  i \hbar \partial_t \psi = \hat{H} \psi
\ee
 corresponding to the hamiltonian
 \be
 \label{2.2} H = {\vec
p^2\over{2m}} \ee with the boundary conditions $\psi(0) =
\psi(L)$. The energy eigenfunctions are given by
\be
 \label{2.3}
\phi_{\vec k} (\vec r) = {1\over{\sqrt{V}}} \exp(- i E t) \exp(i
\vec k . \vec r)
 \ee
  with
  \be
  \label{2.4} V=L^3\ ,\  \vec k =
{2\pi\over L} (n_x, n_y, n_z) \ee
 the $n_i$ being integers. The dispersion
relation reads
 \be
  \label{2.5} E= {\hbar^2\vec k^2\over{2m}}.
  \ee
 The canonical partition function $Z(T,V,N)$ is defined in term of
the hamiltonian  operator $\hat H$ by
\be
 \label{2.6} Z(T,V,N) =
Tr \exp{ \left( - \frac{\hat H}{k T} \right)} .
\ee Using the
complete set of states $\phi_{\vec k}(\vec r)$ displayed earlier,
one finds
\be
 \label{2.7} Z(T,V,1) = \sum_{\vec k} \exp(-{
\hbar^2\over {2m k T}} \vec k^2).
 \ee
 The sum on $\vec k$ is performed on the discrete set given in
Eq(\ref{2.4}). When the box is sufficiently extended and the
temperature not too low, the preceding sum is very well
approximated by an integral. Using $ Z = Z(T,V,N) = (1/N! )
Z(T,V,1)^N$ one then obtains( with a trivial change of the
variable of integration)
 \be
 \label{2.8}
 Z(T,V,N) = {1\over{h^{3N}N!}}
\int d^{3N} p d^{3N} q \exp (-E(q_v,p_v) /kT)
 \ee
 with $ E(p,q)= p^2/2 m $ . The formula $ E(p_v,q_v)$ stands for
 the energy of the system when it is in a configuration in which
 the positions and the momenta of the particles are respectively $
 p_v,q_v$.
  Finally
\be
 \label{2.9}
 Z = {V^N\over {N!}} {1\over{\lambda^{3N}}}
\,\,\,\,\,\,\,\,\,\,\,  \lambda = (\hbar^2/ 2\pi m k T)^{1\over
2}. \ee The Gibbs factor $ N! $ is present only when the particles
are undistinguishable.
 The free energy is related to the partition function by
\be \label{2.10}
 F(T,V,N) = -k T \ln Z .
 \ee

 In the variables $ T,V,N $, the pressure $P$, the entropy $S$ and the chemical potential $\mu$
are given by the relations
 \be
  \label{2.11} P = -{\partial
F\over{\partial V}} \qquad S = -{\partial F\over {\partial T}}
\qquad \mu = {\partial F\over{\partial N}}
 \ee
 while the internal
energy is
\be
\label{2.12} U = F + T S \ee and the constant volume
specific heat   reads
 \be
 \label{2.13} C_V = {\partial U\over
{\partial T}}.
 \ee
 For the case under study,one has
\be \label{2.14}
 \mu = -k T \ln \left\lbrace{V\over N} \left( {2\pi
mkT\over{h^2}} \right)^{3/2}\right\rbrace
\ee
 \be \label{2.15} S =
N k \left[{5\over 2} + \ln \left\lbrace {V\over N} \left({2\pi m k
T \over {h^2}} \right)^{3/2} \right\rbrace\right] \ee \be
\label{2.16}
 C_V = {3\over 2} N k
 \ee
 and the equation of state reads
 \be
\label{2.17} P V = N k T. \ee
\par The ultrarelativistic gas obeys the dispersion relation
$E = c|\vec p|$ .
 The corresponding partition function
\be
 \label{2.18}
 Z_{\ast} = {1\over{N!h^{3N}}} \left({8\pi
k^3T^3\over{c^3}} \right)^N V^N. \ee

 leads to the  equation of state  given in    Eq(\ref{2.17}) and
 the following expressions for the chemical potential, the entropy
 and the specific heat

\be
 \label{2.19}
  \mu = -k T \ln \left[{V\over N} \left(
{(8\pi)^{1/3} hc\over {k T}} \right)^3\right]
 \ee
  \be \label{2.20}
S = N k\left\lbrace 4+\ln \left[{V\over N} \left( {(8\pi)^{1/3}
hc\over {k T}} \right)^3 \right]\right\rbrace
\ee
\be \label{2.21}
C_V = 3Nk. \ee

   The formula Eq(\ref{2.8}) admits a semi classical
   interpretation. The system can classically be symbolized by a
   point evolving in the phase space. The probability for the
   system to be in a configuration in which the first particle is
   in the region $ q_1 \pm \Delta q_1, p_1 \pm \Delta p_1$ , the second particle in
   the region $ q_2 \pm \Delta q_2, p_2 \pm  \Delta p_2, \cdots $  is proportional to $
   e^{(
   - E( \vec{q},\vec{p}))} $ and proportional to the number of elementary cells contained in
   the volume of the
   aforementioned region. At the quantum level, the Heisenberg
   uncertainty relation $ \Delta p_i \Delta q_i \geq \hbar/2  $
   assigns to each elementary cell a volume $ h$. The number of
   such cells contained in the region under consideration is
\be
 \label{2.22}
  \prod \frac{ d^3 p_i d^3 q_i}{ \hbar^3}
   \ee
 which justifies the $ \hbar $ factor in Eq.(\ref{2.8}). The Gibbs
 factor comes from the symmetrization( antisymmetrization) of wave
 functions necessary for the description of a multiparticle bosonic(fermionic)
 state in quantum mechanics. At the semi classical level, it is
 introduced by hand to avoid an increase in entropy when mixing
 two formerly separated quantities of the same gas. We shall see
 in the next section that new commutators  change the
 Heisenberg uncertainty relation and hence the measure needed to
 compute the partition function.

 We
shall mostly interest ourselves to two specific cases in this
work: the relativistic and the non relativistic non interacting
particles. In these two cases,the hamiltonian $H$ solely depends
on the momenta.
 For all the systems we shall consider, we will be interested in the
variations of the thermodynamic quantities induced by the
uncertainty relations coming from the modified commutators. The
new partition functions $Z$ will be related to the usual ones
$Z_{\ast} $ by a relation of the form
\be
 \label{2.23} Z =
Z_{\ast} J^N .
\ee $J$ has to be computed for each modified
commutators. The free energy then becomes
\be
 \label{2.24} F =
F_{\ast} - N k T \ln J.
 \ee
The thermodynamic quantities  are affected in the following way :
\be
 \label{2.25} P = P_{\ast} + N k T {1\over J} {\partial
J\over{\partial V}} \ee \be \label{2.26} S = S_{\ast} + N k \ln J
+ N k T {1\over J} {\partial J\over{\partial T}}
 \ee
 \be \label{2.27}
\mu = \mu_{\ast} - k T \ln J - N k T {1\over J} {\partial
J\over{\partial N}}. \ee $F_{\ast}, P_{\ast}, \cdots$ are the free
energy, the pressure ... obtained from the usual theory, with the
unmodified Heisenberg uncertainty.
 The new internal energy
\be
 \label{2.28} U = U_{\ast} + NkT^2{1\over J} {\partial
J\over{\partial T}} \ee leads to the following  specific heat at
constant volume
 \be
  \label{2.29}
  C_V = C^{\ast}_V + 2 N k T {1\over J}
{\partial J\over{\partial T}} + NkT^2 \left[-{1\over{J^2}}  \left(
{\partial J\over{\partial T}} \right)^2 + {1\over J}
{\partial^2J\over{\partial T^2}}\right]. \ee
 The last system we shall be interested in is a crystal. Two
approaches are available : the Einstein and the Debye models. They
give essentially the same prediction for the specific heat but the
last one, which explains the properties of a solid by advocating
an exchange of phonons between its sites, gives a better
correlation with experiment near zero temperature. We shall
nevertheless use the Einstein model because of its simplicity :
each site is treated as a quantized harmonic oscillator vibrating
at a frequency $\omega$. The partition function is computed
exactly \be \label{2.30} Z(T,V,N) = [2 \sinh
(\hbar\omega/2kT)]^{-N} \ee and leads to the specific heat \be
\label{2.31} C_V = N k \left( {\hbar\omega\over{k T}} \right)^2
{\exp (\hbar \omega/kT)\over {[\exp(\hbar \omega/kT)-1]^2}}
 \ee
which varies from zero to $ 3 R$  for a mole as the temperature is
increased.

\section{K.M.M. theory.}
\par The modified  commutation relations proposed by Kempf,
Mangano and Mann \cite{LABEL1} are of the form \be \label{3.1}
[\hat x_i, \hat p_j] = i\hbar \delta_{ij} (1+\alpha \hat x^2 +
\beta \hat p^2) \ee

 To be more specific, let us consider the case $\alpha = 0$. The
other commutators read
\begin{eqnarray}
[\hat p_i, \hat p_j] &=& 0\nonumber\\
\label{3.2}
[\hat x_i, \hat x_j] &=& 2 i\hbar \beta (\hat p_i \hat x_j - \hat p_j
\hat x_i)
\end{eqnarray}
The last relation means that we have a ''non commutative geometry'' since
 translations along different directions do not commute anymore. The
presence of $\vec p^2,\vec x^2 $ in Eq. (\ref{3.1}) implies that
the rotational  symmetry, whose generators are
\be
\label{3.3}
\hat L_{ij} = {1\over{1+\beta \vec p^2}} (\hat x_i \hat p_j - \hat
x_j \hat p_i)
 \ee
 is  still preserved.
\par One of the remarkable features of this theory is the existence
of a minimal length uncertainty $\hbar \sqrt{\beta}$ because of
the new  Heisenberg inequalities: \be \label{3.4} \Delta x_i
\Delta p_j \geq {\hbar\over 2} \delta_{ij} \left(1+\beta
\sum^n_{k=1} ((\Delta \vec p_k)^2 + <\vec p_k>^2 ) \right). \ee
The momentum representation is found to be given by the operators
\be \label{3.5} \hat p_i .\psi( \vec p) = p_i \psi(\vec p) \ee \be
\label{3.6} \hat x_i . \psi( \vec p) = i\hbar (1+\beta \vec p^2)
\partial_{p^i}\psi(p) \ee acting on a Hilbert space in which the
scalar product is given by \be \label{3.7} <\phi|\psi> =
\int{d^3\vec p\over{(1+\beta \vec p^2)}} \phi^{\ast} (p) \psi(p).
\ee
\par The presence of a minimal length uncertainty implies that no
position representation exists. The concept which proves to be the
closest to it is the quasi-position representation in which the
operators are non local :
\be
 \label{3.8}
  \hat x_i = \xi_i +
\sqrt{\beta} \tan( - \hbar\sqrt{\beta}\partial_ {\xi_i})
\ee
 \be
\label{3.9}
\hat p_i = \left( \sqrt{\beta^{-1}} \right) \tan  ( -i
\hbar \sqrt{\beta} \partial_{\xi_i} ).
\ee

 When we refer to a system as a "relativistic" one , this will obviously
    be linked to its classical behavior. This precision is
    important because
   these deformations  give rise to theories in which even if the the Poincarr\'e group
   was a symmetry at the classical level, it is  broken at the
    quantum level.

\par The constants $\alpha, \beta $ which appear in Eq(\ref{3.1}) are free parameters. What
numerical values can they assume?  Considering the case $ \alpha \neq 0$ it was
suggested \cite{LABEL15} that the minimal length uncertainty $ \hbar \sqrt{\beta} $ should
 be of the order of the Planck scale. We will adopt a more liberal point of view here.
 The only constraint is that  the deformed commutator  must reproduce what is predicted by
 the usual theory and observed experimentally at our energy scale. Our attitude is inspired
 by recent works which have shown that a new physics may take place  well before
 the Planck scale \cite{ LABEL16}. As we shall see in the next sections,
 the behavior of  a non interacting gas can be used  to derive  new, thermodynamic constraints
 on
 these parameters. Typically, the new theories  will predict  new equations of state or (and)
  new specific heats. The experimental values will be used to  obtain  bounds on the parameters
  of the model.

When one introduces the parameters length and momentum scales
$\alpha, \beta$  , it is straightforward that with the Boltzmann
constant $k$ , the light velocity $c$ and the mass $m$ of any
particle, one can construct on purely dimensional grounds the
characteristic temperatures $ 1/( \beta m k) $ and $ c/(k
\sqrt{\beta}) $ ( the first is non relativistic, particle
dependent while the second is relativistic and universal) and the
critical volume $ \alpha^{-3/2}$. We are interested in what
happens at these temperatures and volumes.

\subsection{The case $\alpha = 0$}
\subsubsection{ The non interacting gas }
Let us study a non interacting non relativistic gas. Thinking in
terms of quantum mechanics from the start, one has to solve the
Schr\"Schrodinger equation for a particle in a box. This has to be
done in the quasi position representation when $(\alpha=0)$
because of the lack of a position representation . For simplicity,
let us consider the one dimensional case i.e. Eq(\ref{2.1}):
  with $\hat p$
given in Eq. (\ref{3.9}). One obtains the solution
\be
\label{3.10}
 \psi(t,\xi)\div e^{-iEt} \exp \left(\pm
{i\xi\over{\hbar \sqrt{\beta}}} \arctan \sqrt{2m\beta \hbar E})
\right) .
\ee When the boundary condition
$\psi(t,\xi=0)=\psi(t,\xi=L)$ is imposed, one finds the dispersion
relation
\be
 \label{3.11}
  E_n = {1\over{2m\beta\hbar}} \tan^2
\left( {2\pi  \hbar \sqrt{\beta} n \over L  } \right)
 \ee
 ( n
being an integer) already obtained in \cite{LABEL1,LABEL5}. This
may lead to a cut  off in  order to avoid an infinite energy.
However, this is not obligatory since the energy can become
infinite only for very special values of the parameter $\beta$.If
one does not have a cut-off in the momenta, the energy is no more
an increasing function of the quantum number $n$. If the
deformation parameter assumes a special value which forces one to
consider such a cut-off, then once the number of particles $N$ is
fixed, the energy of the system is bounded.

 In
the two cases, the integration on the momentum which is given
below  has an infinite upper bound. The one particle partition
function
\be
 \label{3.12}
  Z(T,V,1) = \sum^{\infty}_{n=0}
e^{-{E_n\over{kT}}}
 \ee
 can be approximated by an integral if $L$
is big enough. Introducing the integration variable $p$ by $p^2 =
2mE$ and replacing the length by an integral on position, we find
\be
 \label{3.13}
 Z(T,V,1) = {1\over h}\int dx dp {1\over{1+\beta
p^2}} e^{-p^2/2mkT}.
\ee

\par One sees that in the new theory, one can keep the habitual dispersion
relation and just modify the elementary cell volume. This could be
anticipated with a semi classical reasoning. The Heisenberg
uncertainty relation derived from the equation $ [x,p] = i \hbar (
1 + \beta p^2) $ reads
 \be
 \label{3.14}
  \Delta x \Delta p \geq \frac{ \hbar}{2} ( 1+ \beta p^2).
  \ee
It assigns to the elementary cells of the phase space of the new
theory a volume $ \hbar ( 1+ \beta p^2) $ which replaces the usual
value $ \hbar $. From this we find a simple recipe when dealing
with the semi classical approximation:it is obtained by keeping
the classical dispersion relation but modifying the measure in a
way consistent with the Heisenberg uncertainty relation.

\par Let us now go back to the multi dimensional  case. Due to the
presence of three dimensional space dimensions, $\hbar$ is
replaced by $ \hbar^3 $. The rotational symmetry results in the
replacement of $ p $ by $ \vec{p}^2 $ ( This comes from
Eq.(\ref{2.1}) )so that the new partition function reads \be
 \label{3.15}
 Z = {1\over{N! h^{3N}}}\int
{d^{3N} \vec p d^{3N} \vec q\over {(1+\beta \vec p^2)^{3N}}} \exp
(-E(q_v, p_v)/kT).
\ee
\subsubsection{The ultra relativistic non interacting gas.}
   As it is shown in the appendix, the new partition function leads
   to a non relativistic theory which differs very little from the
   undeformed case at habitual temperatures. This could be
   expected since the characteristic temperature $ T_c $ is very
   high for a physically reasonable deformation parameter. The
   only sector where something new is likely to occur is at
   temperatures at least equal to $T_c$. But, at  those temperatures,
   particles are essentially relativistic. With the light
   velocity, the Boltzmann constant and the parameter $ \beta $,
   one can construct, on dimensional grounds, a temperature $
   T_{cr} = c/(k \sqrt{ \beta} ) $ . A new physics will be seen to
   emerge above this characteristic temperature. Working for
   simplicity in the ultrarelativistic limit,
the expression  of $J$ is defined  , similarly to the case treated
in Appendix $ IV.A $ by the integral
 \be
 \label{3.16}
  J = \int_0^{\infty} \phi(x) dx
\ee with
 \be
 \label{3.17}
\phi(x) =  2 \pi^{-1/2}  e^{-x} x^2 \left(1+\beta{k^2T^2\over
{c^2}} x^2 \right)^{-3} \ee with $x={kT\over c}p$.

\par As can be seen from Eq.(\ref{3.17}),  $J$ has no
volume dependence. This saves the relation $ pV=NkT $. The
presence of the temperature and the absence of the number of
particles in the expression of $J$ results in the fact that
 the entropy receives two contributing  terms ( Eq.(\ref{2.20}))    while the chemical
 potential last contributing term vanish ( Eq.(\ref{2.19})).
 The internal energy remains unchanged is modified  and
by way of consequence the specific heat at constant volume .

\par The dominant contribution to the integral is located near the
positive real $ \bar x$  satisfying
 \be
 \label{3.18}
-\beta{k^2T^2\over {c^2}} \bar x^3 - 4\beta {k^2T^2\over {c^2}}
\bar x^2 - \bar x + 2 = 0. \ee
 As explained in the appendix for the non relativistic case, a saddle point approximation is
  easily computed once when
 realizes that
\be
 \label{3.19}
 \phi''(\bar x) = - 2  \pi^{-1/2} e^{-\bar x}
\left(1+\beta{k^2T^2\over{c^2}} \bar x^2 \right)^{-5}.
\ee
 The
relevant zone for the saddle point approximation correspond to
high temperatures. One then finds , when $ T >> T_{cr} $

\be
 \label{3.20}
 \bar x = c/(k  \sqrt{\beta}) \ee Introducing the
universal relativistic characteristic temperature $ T_{c} =  c/(k
T \sqrt{\beta})$, one finds
 \be
 \label{3.21}
 J = e^{ -
\frac{T}{T_{cr}} } \left(\frac{T}{T_{cr}} \right)^3  \frac{1}{
2^{9/4} \pi^{1/4}}
\ee
 \be \label{3.22} S =  N k\left\lbrace 1+\ln
\left[{V\over N} \left( {(8\pi)^{1/3} hc\over {k T}} \right)^3
\left( \frac{T_{cr}}{T} \right)^3 \right]\right\rbrace \ee \be
\label{3.23} \mu = - k T \left\lbrace 1+\ln \left[{V\over N}
\left( {(8\pi)^{1/3} hc\over {k T}} \right)^3 \left(
\frac{T_{cr}}{T} \right)^3 \right]\right\rbrace \ee
  and the equation of state   takes the form
\be
 \label{3.24}
 \rho =  3 p  \left( 1 - \frac{ \sqrt{2}}{2}
\right) . \ee

It is easily found that an adiabatic process takes the form
 \be
\label{3.25}
 V = c^{te} N T^6
\ee
  while the law of action of masses   reads

\be
 \label{3.26}
 \frac{ X_A^ a X_B^b}{X_C^c X_D^d} = c^{te}
T^{-6(a + b -c -d)}
 \ee
   The dependence of the   constants  appearing in the last two equations  on the characteristic
   temperature is obvious.
\par The saddle point approximation used so far gives a picture which, even if it is
qualitatively correct, is not the exact result. For example, when
$ \beta = 0 $, Eq(\ref{3.21}) gives $ J = 0.85$ while one knows
the correct value to be $ J = 1$. The exact result is computed
numerically and plotted in the figures. From them one extracts the
more accurate values of the $ \sigma_i $  and finds that $
\sigma_0 = 1 $ as it should. The presence of the $V,N$ in the
expression of $S$ makes this theory safe concerning the mixing
entropy and the the second law of thermodynamics, contrary to what
happens in the case $ \beta = 0 $ as we shall see in the next
section . The equation of state takes the form
 \be
 \label{3.27}
\rho = 3 p ( 1 + g(T/T_{cr})) \ee
   The properties of the function $ g(x)$  are obtained numerically. In Fig.1 is
   displayed  the  ratio of the new energy density  and the habitual one,
     as a function of the variable $ T/T_{cr}$.
\par  Let us now give an estimate of the bound thermodynamics  imposes on the
deformation parameters. Consider the Helium   whose specific heat
at constant volume assumes the experimental values  $12.4  J
K^{-1} mole^{-1}$ (The undeformed theory assigns the value $12.47$
to any non relativistic  gas). With the help of the appendix, we
can compute the specific heat in the new theory $ C_V =  12.47(1+
\sigma( \beta m k T))$. The measured value tells us that $ 12.39 <
C_V < 12.59$. At first order, $ \sigma(x) \sim x^2 $. Assuming $T
= 300^o K$, this gives $ \beta \geq 10^{- 45}$. The unit in which
$ \ beta $ is given is $ s^2 kg^{-2} m^{-}$. One then finds
 a minimal length uncertainty $ \gamma \geq 10^{-16} $ meters  which is quite close to
 the bound derived from atomic physics considerations \cite{LABEL12}. Working with another
  inert gas( the argon), we end up with  the same result. This value of $ \beta $ can now be
  used to obtain
  a bound on  the  relativistic characteristic
  temperature,  $ T_{cr} \geq 10^{9} K$. This is very far from the
  value $  T_{cr} \sim 10^{16} K$ which corresponds to a minimal  uncertainty in
  position of the order of the Planck scale.
  Taking the modified relations used
  here as a "fundamental theory" i.e. not a phenomenological approximation( coming from string
  theory for example) , one sees that a new physics may appear well
  before the Planck scale. Its specific effects  will be seen essentially in the early
  universe
   but astrophysics may also provide a field of application since the temperatures involved can
   be near the characteristic relativistic temperature whose lower bound is $ 10^9 °K $.
   The case  $ \beta =0$ has been seen to have a specific heat which is constant
    ( see $ Eq.(\ref{2.29}) $ ); the argument
   we have  used with the argon and the nitrogen to infer a bound on $ \beta $ in the case
   $ \beta \not 0 $ , can not be repeated here to constrain $ \alpha
   $. One can  use the departure of the equation $ Eq.(\ref{A.15}) $
   from the relation $ P V = N k T $ to find the allowed zone
    $ \alpha \leq 10^(-3)$ ( the unit is $ m^{-2} $ ) which is quite
    large  .

\subsubsection{The Einstein model for a solid.}
 When $\alpha = 0$,
the spectrum of a harmonic oscillator is found to display a
quadratic term \cite{LABEL1} :
 \be
 \label{3.28}
 E_n = E_0 + an + b
n^2
 \ee
 with
 \be
  \label{3.29}
  E_0 = \frac{1}{2} \hbar \omega
\left( \frac{1}{4 \sqrt{r}} + \sqrt{1 + \frac{1}{16 r}} \right)
\ee
 \be
  \label{3.30} a =  \hbar \omega \left( \frac{1}{4 \sqrt{r}}
+ \sqrt{1 + \frac{1}{16 r}} \right)
 \ee \be \label{3.31} b =
\frac{ \hbar \omega}{4 \sqrt{r}}
 \ee
 where
\be \label{3.32}
   r = \frac{1}{(2 m \hbar \omega \beta)^2}
\ee
so that $b$ vanishes when $\beta \rightarrow 0$.
 The free energy of a crystal with $N$ such oscillators at its sites
is
\be
 \label{3.33}
 F = N(E_0-k T \ln K_0(T,\beta))
 \ee with
  \be
\label{3.34} K_0(T,\beta) =\sum^{\infty}_{n=0}\exp \left(-{an+b
n^2\over {k T}} \right). \ee
 Its specific heat at constant volume reads
\be
 \label{3.35}
 C_v = 3Nk\left[2T
{K_1(T,\beta)\over{K_0(T,\beta)}} +
T^2{K_2(T,\beta)\over{K_0(T,\beta)}}
-T^2\left({K_1(B,\beta)\over{K_0(T,\beta)}}\right)^2\right] \ee
with
\be
\label{3.36} K_1 = {\partial K_0\over{\partial T}} \ and\
K_2 = {\partial K_1\over {\partial T}} . \ee
  A numerical treatment shows that the behavior of the specific heat in this model is
   very close to the one obtained in the undeformed case, for values of $ \beta $ in the
   range we found for the non interacting gas in the previous section. There appears to be a
   difference  near
   the zero temperature  but it is so tiny that it is experimentally undetectable
.
\subsection{ The case  $ \beta = 0$.}

\par  In the new context, the Schr\"{o}Schrodinger equation
in the position representation leads to the energy spectrum \be
\label{3.37}
  E_n = \frac{1}{2 m} \left( \frac{2 \pi \hbar \sqrt{\alpha}}{ \arctan{\sqrt{\alpha} L}}
  n  \right)^2
\ee
  for non relativistic non interacting particles in a uni
dimensional box of length $L$. This leads to a measure which could
be inferred from
 the uncertainty relation
\be
 \label{3.38}
  \Delta q \Delta p  \geq {\hbar\over 2} (1+
\alpha q^2). \ee The partition function corresponding to the three
dimensional case is
 \be \label{3.39}
 Z = {1\over {h^{3N} N!}} \int {d^{3N}\vec p
d^{3N}\vec q\over{(1+\alpha\vec q^2)^{3N}}}\exp (-E/kT).
 \ee
 The measure
of the phase space  displayed in Eq.(\ref{3.39}) lowers the
influence of the large $ \vec{q}^2$ i.e of the boundaries .

\subsubsection{Non interacting non relativistic gas.}

\par   In this case the quantity $J$
defined in Eq. (\ref{2.23}) is found to be
 the volume integral
\be \label{3.40} J = {1\over V} \int {d^3\vec q\over{(1+\alpha
\vec q^2)^3}}. \ee
\par As can be seen from Eq.(\ref{2.25}), the fact that $J$ has an explicit
volume dependence breaks the relation $ pV=NkT $. The last
contributions to the entropy and the chemical potential vanish
also because $J(\alpha, V)$ doesn't depend on the temperature and
the particle number. The internal energy remains unchanged ( $ U_*
= U $ ) and by way of consequence the specific heat at constant
volume is also unmodified.
\par   Usually, the positions appear in the integrand of the partition function through
interaction terms. The interaction between two particles depends
on the vector separating them but this is not the case of
Eq(\ref{3.12}). The situation may look similar to that of  a
perfect gas in a constant gravitational  field since the particles
do not interact among themselves
 but couple to an external field. In this case, the partition function is the integral of the
 quantity $ \exp{ ( - \vec{p}^2 - mg z)/(k T)} $. One sees that for  the case  , the factor
 $ (1+\alpha \vec{q}^2)$ lacks the
 proper $ k T$ piece to allow such a parallelism. { \em As a consequence, the thermodynamics
  generated by the K.M.M theory
  can not be interpreted as the usual one with a non trivial interaction}. We will see that
  the same consideration applies to the $q$ oscillator in the last section.

\par The appearance  of the norm of the position vector raises the question of the origin
 from which it is measured. This is related to the question of frame dependence which is not
  specific
  to the theories under study. In habitual statistical physics, the partition function involves
  the energy which is not the same for all observers, even galilean. The frame chosen is
  that of an observer at rest with respect to the container. In
  the case we are studying here, one has to specify a special
  point which will play the role of the origin. If for example we
  consider a spherical container, the center is a  point with special status.

\par The integral displayed in Eq(\ref{3.40}) is easily performed if we put the gas in a
spherical container of radius $R$ : \be \label{3.41} J = {1\over
V} { \pi \over {\alpha}} \left[-{R\over {(1+\alpha R^2)^2}}
+{1\over 2} {R\over {1+\alpha R^2}} + {1\over {2\sqrt{\alpha}}}
\arctan (\sqrt{\alpha} R)\right]. \ee The Eqs. , (\ref{2.25}),
(\ref{2.26}), (\ref{2.27})show that the pressure, the entropy and
the chemical potential are all affected by the new commutation
relations. Let us, for simplicity, study two limiting cases : the
small and the huge volumes.

\par What happens when the volume can not be neglected in front of
$ \alpha^{-(3/2)} $? The expression of $J$ becomes \be
\label{3.42} J = {1\over{\alpha^{3/2}V}} \left[ a - b
{1\over{\alpha^{3/2}V}} + c {1\over{\alpha^{5/2}V^{5/3}}}\right]
\ee
with
 \be
  \label{3.43} a = 2.4674, b = 17.546, c = 78.6488.
 \ee
 Retaining the first two
contributions, the pressure takes the form \be
 \label{3.44} P
= NkT\left({b\over a} {1\over{\alpha^{3/2}}} {1\over{V^2}} -
{11\over 3} {c\over a} {1\over
{\alpha^{5/2}}}{1\over{V^{2/3}}}\right).
\ee

\par  This theory predicts an important change in the thermodynamic
properties for huge volumes; the correction to the pressure for
small volumes given  in Eq. (\ref{A.15}) in the appendix  is
negligible. But the last expression shows that for very big
containers, the pressure of a non relativistic gas may become
negative. However, this does not contradict any experimental bound
since for  typical values of $ \alpha  $ like the ones of
\cite{LABEL12}, \cite{LABEL15} , one would need a recipient  whose
volume is greater than the earth . A negative pressure is a priori
puzzling because it does not appear in everyday thermodynamics.
But one knows it is an essential ingredient of the inflation
scenario in cosmology \cite{LABEL17}. We will come back to this
point in the conclusions.

Similarly, one finds for the entropy and the chemical potential in
the case of important volumes :
\be
\label{3.45}
S =
Nk\left\lbrace {5\over 2}+\ln \left\lbrace
  \left({a\over{\alpha^{3/2}}} \right) {1\over N} \left({2\pi mkT\over{h^2}} \right)^{3/2}\right\rbrace \right\rbrace
\ee
 \be
 \label{3.46} \mu  = -kT\left\lbrace    \ln \left\lbrace
 \left( {a\over{\alpha^{3/2}}} \right) {1\over N} \left({2\pi mkT\over{h^2}} \right)^{3/2}\right\rbrace \right\rbrace.
\ee
\par The formula giving the entropy in big containers does not involve the volume itself and
 tells us that  an adiabatic expansion is nothing else than a isotherm. The expression of the
 chemical potential shows that when the reaction
\be \label{3.47}
  a A + b B \longrightarrow c C + d D
\ee
  attains equilibrium, the following relation holds between the number of particles
   $N_i$ of the species $i$

\be \label{3.48}
 \frac{ N_C^c N_D^d}{N_A^a N_B^b} =  \tau T^{3/2 (c + d - a - b)}
\ee
   with $ \tau$ a quantity which does not depend on the temperature or the pressure. This is
   very different from the action of masses law in which only the concentrations  are
   constrained by a similar relation.

  We now investigate entropy mixing. Consider two gases $ A,B$   contained in two
  compartments ( of respective volumes $V_A, V_B$ of a container separated by an
  impermeable wall). They are at the same temperature and pressure, their respective numbers of particles are $N_A,N_B$ .Now, let us remove the wall. In the non deformed theory($ \alpha = \beta = 0$), the temperature and the the pressure remain the same ( due to the conservation of  energy and the equation of state). According to Eq.(\ref{3.22}), the increase of entropy for this process
is non vanishing.

   In the theory we are studying, the equation of state specified by Eq.(\ref{3.44}) and the
   internal energy lead to the same pressure and temperature after mixing. However,
   the expression of the entropy is now given by Eq.(\ref{3.45}) .
   For huge volumes, Eq.(\ref{3.22})  gives
that $ \Delta S $ vanishes.
   This  differs significantly from the non deformed theory. Let us consider two important but
   equal volumes $ V_A = V_B =V$ and two equal number of particles $ N_A = N_B = N$ of two
   different gases. The habitual theory states that the entropy mixing is $ 2 N k \log{2} $
   while this K.M.M deformation predicts $ \Delta S = 0 $.

\subsubsection{The ultra relativistic non interacting gas.}

$J$ is found to be exactly the same as in the   previous section
so that compared to the undeformed case, the state equation, the
entropy and the chemical potential change while the specific heat
at constant volume remains the same.
 \be
  \label{3.49} S =
Nk\left\lbrace    4 + \ln  \left[  \left({a\over{\alpha^{3/2}}}
\right) {1\over N} \left({(8\pi)^{1/3}hc\over {k T}}
\right)^{3}\right]\right\rbrace
 \ee
 \be \label{3.50} \mu = -k T  \ln
\left\lbrace   \left({a\over{\alpha^{3/2}}} \right) {1\over N}
\left({(8\pi)^{1/3}hc\over {k T}} \right)^{3/4} \right\rbrace
 \ee
 The absolute value of the
chemical potential is lowered compared to the habitual case. The
formula Eq(\ref{3.25}) is still valid, but with a different $
\tau$. In summary, above the critical volume $ \alpha^{- 3/2}$,
the equation of state takes the form
 \be
  \label{3.51}
 \frac{P V}{N k T} = 1 + f( V/V_{cr}) \qquad  \rho = 3 p (  1 + f( V/V_{cr}))
\ee
 for a relativistic gas; in the non relativistic case, the
factor $3$ in the last equation should be divided by $2$. The
function $ f(x)$ is plotted in Fig.2 . It begins at zero for a
vanishing $x$ and quickly hugs to the horizontal asymptot $ y = -
1$. In this model, the parameter $ \alpha $ is fixed once and for
all. If one considers an increasing volume for a fixed quantity of
gas( fixed $N$) and a fixed temperature( fixed $T$), the last two
equations show that $ \rho/p $ and $ p/T $ go to zero.

\par  An important  issue for gases presenting a  non trivial equation of state is the
 possibility of phase transitions. For example, a Van der Waals gas
 exhibits a
  condensation plateau when  it is  compressed at (fixed temperature )under a critical volume
   ( obtained by Maxwell
  construction). In the phase diagram , this is linked to the existence of a critical
  point where the two  first derivatives of the pressure with respect to the
  volume( on the appropriate isothermal curve) are
  vanishing.
 From Eq.(\ref{3.30}) one can isolate $P$ and see that
\be
 \label{3.52}
 \frac{ \partial P}{\partial V} = \frac{N k
T}{V^2} f_1(V/V_{cr})   \qquad \frac{ \partial^2 P}{\partial V^2}
= \frac{N k T}{V^3} f_2(V/V_{cr})
\ee

\be \label{3.53} f_1(x) = - 1- f(x) + x f'(x) \qquad f_2(x) = 2 +
2 f(x) - 2 x f'(x) + x^2 f''(x) \ee The functions $f_1(x)$ and
$f_2(x)$ are plotted in Fig.3 an Fig.4 respectively. They show
that there is no such a phase transition here.

\par   What we have done for gases  is similar to Maxwell Boltzmann
statistics because we did not symmetrize the wave functions for
the bosons nor antisymmetrize for the fermions. At the fundamental
level, one can ask if the concept of spin is relevant in these
theories and, in the case the answer is positive, one may wonder
about the spin-statistics theorem in the new context.
\par   The commutation relations given in Eq. (\ref{3.1}) and Eq.(\ref{3.2}) are not
invariant under a Lorentz transformation. As the spin of a
particle is defined through the behavior of its wave function
under this group, one apparently looses the notion of spin here.
However, this is not obligatory true. For example, taking the
ultimate structure of space-time to be a particular commutative
geometry, the relevant   group of isometries  is not the
Poincarr\'e one but a $q$ deformation of it. There is a new
algebra   but a spin has been defined in these theories and the
wave functions for particles of spin $1/2, 1$, etc... have been
found \cite{LABEL18}. Although the question has not been addressed
in K.M.M. theory, we expect a similar situation to occur.
Considering as an illustration the replacement of the Dirac
equation by
\be
 \label{3.54}
 (\gamma^{\mu} \hat p_{\mu}+ m)\psi =
0 \ee with $\hat p_{\mu}$ given by the generalization  in three
dimensions of Eq. (\ref{3.9}), the relevant symmetries of this
equation should be linked to an inhomogeneous algebra since the
operator is non-linear. We expect a generalization  of the notion
of spin which may conserve the spin-statistic theorem. This is not
really crucial here since we are performing a Boltzmann like
construction which is known to be valid at high temperature and
does not discriminate between bosons and fermions. What we called
" relativistic systems" in the deformed theories  are nothing than
systems described by the habitual relativistic equations, but with
the momenta replaced by their quasi position expression in the
model where $ \alpha $ vanishes, for example. This can in
principle be made rigorous using the functional integral
introduced in \cite{LABEL3} since in this formalism one does not
need to introduce neither the commutation relation nor the scalar
product from the start.
\par The theories studied here are  unusual in many respects. we have seen that a minimal
length uncertainty can exist for specific parameters. The fact
that the commutator of two quantum operators  is not simply
proportional  to the Poisson bracket of their classical
counterparts creates a situation in which a quantity conserved at
the classical level may not be conserved at the quantum level.
This implies that one has to be careful when studying ensembles
more complicated than the canonical one.
 \par  We studied a system which exchanges only energy with the surroundings. When the number
  of particles is not fixed, one has to work in the grand canonical ensemble. However, it is
  known in usual thermodynamics that the two descriptions yield essentially the same
  predictions( equation of state, specific heat, $\cdots $). A real difference appears
  essentially when describing phase transitions.

\par Our treatment concerned only non interacting gases but we know they only approximate
 the real ones when the density is small. We think it is    physically irrelevant to include
 interactions for habitual gases like hydrogen, ammonia, nitrogen, etc $ \cdots $ . Taking for example
  the theory where only $ \beta \neq 0$,
we saw that a new behavior sets in at very high temperature. But
at this energy scale, nitrogen does not exist anymore and the use
of an interaction of the Lennar-Jones type is not justified; at
very high temperatures, the existing species are elementary
particles like photons, neutrino, leptons, quarks $ \cdots $. One
may expect to have a different situation for the model in which
only $ \alpha $ is non vanishing ; one may conceive that the
expansion of the universe may provide the necessary huge volumes.
However, as the universe gets bigger its density decreases and the
mean distance between the particles becomes too important ;they
nearly don't feel the interaction.

\section{The $q$ deformed oscillator.}

 A space with non commutative coordinates is called a quantum
plane; its covariant differential calculus has been investigated
\cite{LABEL19}. A quantum group can be understood as a set of linear
transformations acting on it. The commutation relations compatible with
this structure \cite{LABEL20} are quickly summarized below.
\par The momentum  and position  operators $ P,X$ satisfy the
commutation relation
\be
\label{4.1}
 q^{1/2} X P - q^{-1/2} P X  =  i U.
\ee
The deformation parameter $q$ is strictly greater than one; the unitary
element $U$ has been introduced to allow hermitian momentum and position :
\be
\label{4.2}
 P^+ = P \,\,\,  X^+ = X \,\,\,  U^+= U^{-1} .
\ee
It is also subject to the conditions
\be
\label{4.3}
U P = q P U \ ,\  U X = q^{-1} X U.
\ee
Introducing the extra parameter $ M $ , one defines the operators
\be
\label{4.4}
a^+ = \bar \alpha U^{2M} + \bar \beta P U^M
\ee
\be
\label{4.5}
a = \alpha U^{2M} + \beta U^{-M} P
\ee
and finds that they obey a modified Heisenberg algebra
\be
\label{4.6}
aa^+-q^{-2M}a^+a = 1.
\ee
\par We first investigate what changes in the Einstein model of a solid
.
In this theory, the harmonic oscillator has been defined \cite{LABEL16}
as having the hamiltonian
\be
\label{4.7}
H=\omega a^+a
\ee
with $a^+,a$ the operators defined in Eqs.(\ref{4.4}),(\ref{4.5}) and whose
algebra is given by Eq. (\ref{4.6}).
\par A special role is played by the eigenvalue
\be
\label{4.8}
E_{\infty} = {\omega\over{1-q^{-2M}}}
\ee
of the hamiltonian :
\begin{description}
\item [a)] There is an infinite subset of the spectrum which is bounded
from above by $E_{[\infty]}$ :
\be
\label{4.9}
E^{\ast}_m = \omega {1-q^{-2M m }\over{1-q^{-2 M}}} \leq E_{[\infty]}
\qquad m = 0, 1,2,\cdots .
\ee
For these states, $a$ raises the energy while $a^+$ lowers it.
\item [b)] The remaining part of the spectrum
\be
\label{4.10}
E_n = \omega \left( {1\over{1- q^{2 M}}} + \beta \bar \beta q^{2n} \right) \,\,\,
n = -\infty,\cdots,+\infty
\ee
(with $\beta$ depending on the mass $m$ and the frequency $\omega$) faces
the reverse, common situation with $a^+$ playing the role of a creation
operator.
\item [c)] For $E = E[\infty]$, $a$ and $a^+$ do not change the
eigenvalue.
\end{description}
To give an example, the operators
\be
\label{4.11}
\hat P = -i{\partial\over{\partial x}} +{1\over{\sqrt{1-q^{-2M}}}}
\gamma\ ,\ \hat X= x
\ee
\be
\label{4.12}
U = q^{-{i\over 2}(\hat X \hat P + \hat P \hat X)}
\ee
are found to obey the relations given in Eqs. (\ref{4.1}-{4.3}) which
define the $q$ deformed phase space. Introducing the operators $a,a^+$ as
specified by Eq. (\ref{4.4}-\ref{4.5}) and expanding the hamiltonian
$\omega a^+ a$, one finds it consists of two terms. The first
part is the hamiltonian of a free oscillator and the remaining part
can be seen as an interacting term 0 as a series in $\sqrt{h}$
with $h = e^q-1$.
\par The link between the parameters $\alpha,\beta,\gamma$ on one
hand and the mass, the frequency of the unperturbed oscillator on the
other hand are
\be
\label{4.13}
\alpha = \pm {i\over{\sqrt{2Mh}}} \qquad \gamma = \pm \sqrt{2m\omega}
\qquad \beta = {i\over{\sqrt{2m\omega}}}.
\ee
\par It should be emphasized that the K.M.M. oscillator presents a
spectrum which reduces to the habitual one as the deformation
parameter is sent to zero. This is evident from Eqs. (\ref{3.29})
and (\ref{3.30}). The situation of the $q$ algebra is different
and so one expects it should lead to an entirely new
thermodynamics, even for a small value of the deformation
parameter.  Let us study a crystal whose sites are occupied by $q$
oscillators.
\par Constructing the partition function
\be \label{4.14} Z(T,V,1) = \sum^{\infty}_{m=0}
e^{-{E^{\ast}_m\over{kT}}} + \sum^{+\infty}_{n=-\infty}
e^{-{E_n\over{kT}}} \ee one finds that the first sum diverges
since $E^{\ast}_m < E_{[\infty]}$. How can we handle this? The
extreme reaction would be to disregard the all theory as
unphysical. Another possibility may be a renormalization of the
partition function but this is questionable since we are not doing
quantum field theory. Another point of view is that these states
may be interpretated as  spurious and  eliminated. We do not have
a clear option between these possibilities for the moment.
Introducing $\sigma = \sqrt{\omega\beta \bar \beta q}$ one
rewrites the remaining part as \be \label{4.15} Z(T,V,N) =
e^{-{\omega\over{(1-q^2}) k T}} \sum^{+\infty}_{n=-\infty} e^{-
\frac{\sigma^n}{k T}}. \ee If $\sigma <1$, one has \be
\lim_{n\rightarrow+\infty} \exp(-\sigma^n) = 1 \ee so that
$Z(T,V,1)$ still diverges. The last option would lead us  to drop
also the states corresponding to high positive values of $n$. The
question would then be  to fix the index where the truncature
should take place.
\par If we discard the $E^{\ast}_m$, we were left with $E_{[\infty]}$
as the lowest state. As it corresponds to $E_n$ for $n=0$, we think it
should be kept in the physical spectrum. So, our truncature begins
at $n=1$ in this case :
\be
\label{4.16}
Z(T,V,1) = e^{-{\omega\over{(1-q^2) k T}}} \sum^0_{n=-\infty}.
e^{- \frac{\sigma^n}{k T} }
\ee
\par When $\sigma>1$, the situation is reversed :
\be
\label{4.17}
Z(T,V,1) = e^{-{\omega\over{(1-q^2) k T }}}\sum^{+\infty}_{n=0}
e^{ - \sigma \frac{q^n}{k T}}.
\ee
  Writing $ q = 1 + h $ and retaining the first two terms of the binomial formula to
  approximate $ q^n $, one  deals with a partition function similar to the one used in the
  K.M.M model. The same results apply, provided one chooses a value of $h$ in the range which
  gives to $ \beta $ the values compatible with the bound given  in section
  $3$. Discarding the extra states has one nice feature: the
  thermodynamics is  smooth in the deformation parameter.

\section{Conclusions}
\par Our investigation of the thermodynamics implied by two special
K.M.M. deformations of the commutation relations have shown that a
new physics appears at high temperatures in one case and for huge
volumes in the other one. In the two cases, the universe should
provide the best laboratory since the modified commutators can
produce specific, visible signatures . In fact, the evolution of
the universe will be affected in more than one aspect. Of course,
the new  equation of state of the radiation gas which fills the
early universe influences the evolution of the scale factor
through Einstein equations. In addition, the new expressions of
the entropy change the form of the adiabatic processes. As in
standard cosmology the expansion of the universe is thought to be
isentropic, this will enter into play. The fact that chemical
potentials receive contribution corrections will affect the
densities at equilibrium . Finally, the theory exhibiting a
minimal length uncertainty may forbid the singularity present in
the standard  big bang scenario. A similar reasoning was advocated
to argue that the Hawking evaporation of a black hole may halt
without using the complementarity hypothesis \cite{LABEL6}.

Let us say a few words about the negative pressures found in section{\ref{3.3}} . In the
inflation scenario, one uses a scalar field whose classical time evolution generates such a
non positive pressure for a given period, because of a well chosen
 potential \cite{LABEL17}. Eq(\ref{3.21}) suggests that one may achieve such pressures
 with ordinary matter, provided the commutation relations are modified. But to achieve
 inflation, one needs the relation $ p < - \rho/3 $ and this not obvious from Eq.(\ref{3.21}).
  One also needs a mechanism to return to normal matter i.e positive pressures.
  Certain mechanisms of inflation are inefficient because their period of exponential
  growth
  never ends. This is the case in topological inflation where the core of a monopole,
  a cosmic string or a domain wall expands endlessly. It should also be stressed that
  here the  negative pressures appear only when the volume gets big enough, contrary to
  standard
  inflation where it happens very early in the course of the
  universe.But the true challenge here, which may rule out all the
  model corresponding to the case $ \beta = 0 ,\alpha \neq 0$ is
  that there is no evidence for a preferred origin. A straightforward possibility is that this theory
may be interpreted as imposing a maximal physical  volume since
going above it one gets into trouble.  One of the most important
novelties    is the behavior of the  entropy mixing  in the $
\alpha = 0 $ case which seems to spoil the interpretation of the
entropy derived from Eq. (\ref{2.11}) as a variable counting the
number of states. The breaking of homogeneity makes it likely that
one should consider Tolman (position dependent) temperatures like
in curved space times. But in such spaces the fact that the
quantity $ T \sqrt{- g_{00}}$ is constant is a vital information
which seems to have no counterpart in the theory $ \alpha \not = 0
$. This model is also useless in cosmology since the existence of
an origin is in contradiction with the homogeneity hypothesis. One
may use all
  the undesirable consequences of this the model to argue it is
  unphysical, like the quantum theory on a Moyal plane modifying
  the time-position commutator $ [ x_o,x_i]$ has been discarded
  because it leads to a non unitary theory.
\par We plan to analyze the cosmological incidences of the model $ \alpha = 0$ in a near future.
The influence
  of the new physics
in the early universe is at the center of a recent work using
Unruh and  Jacobson modified dispersion relations \cite{LABEL21}.
Our approach will differ by the fact the dispersion relation will
come from an assumption on the quantum space time . The
commutation relations studied here can be interpreted  as
phenomenological consequences of string  or M theory
\cite{LABEL22, LABEL23}. This paper suggests that string cosmology
is not uniquely characterized by the evolution of the fields which
appear at the lowest order i.e. dilaton, $\cdots $  but also by
some non trivial statistical effects. \cite{LABEL17}.

\par  We have restricted ourselves to the two extreme cases $ \alpha = 0 $ and $ \beta = 0 $.
One can ask what happens when $ \alpha $ and $ \beta $ are
simultaneously non vanishing. An uncertainty relation similar to
the one displayed in Eq. (\ref{3.4}) exists but one does not
possess a position nor a momentum representation. To deduce
classical statistical physics from quantum statistical physics as
we did  , one has to deal with the Bargmann representation
\cite{LABEL3} which is more complicated. In this case , $J$ will
depend on both $ T$ and $V$. The reserves we expressed due to the
presence of a preferred origin also apply here.
 \par Can a similar treatment be performed for gases in the $q$
deformed algebra? The situation may prove difficult since for a free
particle in this theory, the momentum is quantized without advocating
boundary conditions \cite{LABEL10}.

\underline{Acknowledgement} It is a pleasure to thank Ph.Spindel
and R.Brout for stimulating discussions.

\section{Appendix}

\subsection{The Non relativistic non interacting gas in the $
\alpha = 0 $ theory}

  The change of variable $x=\vec p^2/2mkT$, allows one to write
 $J$ as the dimensionless integral
\be
 \label{A.1}
 J = 2 \pi^{-1/2} \int^{\infty}_0 dx e^{-x}
x^{1/2} (1+2\beta m k T x )^{-3}. \ee
 This integral can not be
performed analytically. Nevertheless a useful approximation can be
obtained which gives a good qualitative picture. Being the product
of an increasing and a decreasing exponential, the integrand
presents a maximum at
\be
 \label{A.2}
 \bar x = {-(1+5\beta m k T) +
\sqrt{1+14\beta m k T+25 \beta^2m^2k^2T^2} \over{ 4\beta m k T}}.
 \ee
Calling $\phi(x)$ the integrand in Eq.(\ref{3.34}) , a saddle
point approximation gives
\be
 \label{A.3}
 J = (\phi( \bar
x))^{3/2} {\sqrt{2\pi}\over{\sqrt{-\phi''(\bar x)}}}.
\ee
The fact
that $\phi(x)$ is maximal at $\bar x$ simplifies the expression of
the second derivative
\be
 \label{A.4}
 \phi''(\bar x) = (1+4\beta
m k T \bar x) (4 \beta m k T \bar x^2-4\bar x + 1). \ee
 In this case
$J$ is a function of  the temperature
 but it does not depend on the particle number or the volume. Consequently the equation
  of state
$PV = NkT$ is  still valid here but the molar specific heat at
fixed volume is no more the constant $3 R/2 $. Let us compute the
first correction to $J$. A gas presents a non relativistic
behavior at low temperature; in this limit we obtain, introducing
the particle dependent characteristic temperature $ T_{c} = 1/
\beta m k $
 and find to the first order in $ T_{c}/T $
\be
\label{A.5}
 \bar x = \frac{1}{2} - 3 \frac{ T}{T_{c}}
 \ee \be
\label{A.6} J =  \sigma_0 + \sum_l  \sigma_l ( T/T_{c})^l \ee
 with

\be
 \label{A.7}
   \sigma_0 = 1.20353    , \sigma_1 =  - 10.8318,  \sigma_2 = 93.8755 ,  \cdots.
\ee

Now
 \be
 \label{A.8} S = N k \left[{5\over 2} + \ln{ \sigma_1} + \ln
\left\lbrace {V\over N} \left( {2\pi m k T \over {h^2}}
\right)^{3/2}    \right\rbrace + 2  \frac{\sigma_1 }{\sigma_0}
\left(\frac{T_{c}}{T} \right)
 \right]
\ee

\be
\label{A.9}
 \mu = - k T  \left[ - \ln{ \sigma_0} + \ln
\left\lbrace {V\over N} \left( {2\pi m k T \over {h^2}}
\right)^{3/2}
  \right\rbrace  +  \frac{\sigma_1  }{\sigma_0}  \frac{T}{T_{c}}   \right]
\ee
   and the equation of state is
\be
 \label{A.10}
 \rho = P \left( \frac{3}{2} - \frac{\sigma_1}{\sigma_0} \frac{T}{T_{c}}. \right)
\ee
   An adiabatic process takes the form
\be \label{A.11}
 V = c^{te} N T^{-3/2} e^{  \frac{ 2  \sigma_1}{\sigma_0}  \frac{T}{T_{c}} }
\ee
   while densities $ X_i = N_i/V$ at equilibrium for the reaction  displayed in Eq.(\ref{3.24})
    obey
\be \label{A.12}
  \frac{ X_C^c X_D^d}{X_A^a X_B^b} = c^{te} T^{3/2( -a-b+c+d)}
e^{ \left( - \frac{\sigma_1}{\sigma_0} ( \frac{a}{T_{c}^A} +
\frac{ b}{T_{c}^B} + \frac{c}{T_{c}^C} +\frac{ d}{T_{c}^D})
\right)} \ee
  since each particle, having its own mass, possesses a specific critical temperature.

\subsection{The Non relativistic non interacting gas in the $
\beta = 0 $ theory }
 If one considers a small container, then the following
approximation is accurate for the quantity $J$ :
\be
 \label{A.13}
J = 1-a_1 \alpha V^{2/3} \ee with
\be
 \label{A.14} a_1 =
{4\over{5}} \left( {3\over 4} \right)^{5/3} \pi^{-2/3} .
 \ee
 The
new pressure
 \be
  \label{A.15} P = N k T V^{-1} \left( 1-{2\over 3}
a_1 \alpha V^{2/3} \right)
\ee
is lower than the one obtained for
the same gas in the unmodified algebra. The minus sign suggests
that the pressure may become negative. However, as $\alpha$ must
be tiny, this would occur for huge volumes but then the
approximation specified in Eq. (\ref {3.14}) is no more valid.
\par The entropy and the chemical potential become
\be
 \label{A.16}
 S = Nk\left[{5\over 2} - a_1 \alpha V^{2/3} + \ln
\left\lbrace {V\over N} \left( {2\pi mkT\over{h^2}}
\right)^{3/2}\right\rbrace \right]
 \ee
 \be
  \label{A.17}
  \mu =
-kT\left[\ln \left\lbrace {V\over N} \left( {2\pi mkT\over{h^2}}
\right)^{3/2} \right\rbrace + a_1\alpha V^{2/3}\right]
\ee
 while
the specific heat remains the same. One sees that for small
volumes nothing really new appears.

\section{figure captions}
\begin{itemize}
\item Fig1 The quantity $ 1 + g(T/T_{cr})$ shows that
      the energy density due to a radiation  of fixed pressure
      decreases at high temperatures in the $ \alpha = 0 $ case.
\item Fig2 The function $ 1+ f(x)$ which modifies the equation of state when $ \beta = 0 $
         is depicted. It tends quickly to its asymptotic value.
         For a system with fixed $T,N$, the pressure falls more
         quickly when the volume is increased in comparison to the
         undeformed case.
\item Fig3 The function $ f_1(x)$ linked to the first derivative of the
      pressure does not vanish for finite volumes so that the
      isotherms in the $ \beta = 0 $ theory do not have an extremum in
      the $ P,V $ diagram.
\item Fig4 The function $ f_2(x)$ linked to the second derivative of the pressure does
      not vanish neither. This and the previous figure proves
      there is no phase transition.

\end{itemize}


\begin{thebibliography}{18}
\bibitem{LABEL1}
A. Kempf, G. Mangano, R.B. Mann, Phys. Rev. D {\bf{52}}, 1108 (1995).
\bibitem{LABEL2}
M. Hinricksen, A. Kempf, J. Math. Phys. {\bf{37}}, 2121 (1996).
\bibitem{LABEL3}
A. Kempf, J. Phys. A : Math. Gen. {\bf{30}}, 2093 (1997).
\bibitem{LABEL4}
A. Kempf, G. Mangano, Phys. Rev. D {\bf{55}}, 7909 (1997).
\bibitem{LABEL5}
R.Brout, C.Gabriel, M.Lubo and Ph.Spindel, Phys.Rev.{\bf{D59}}
044005 (1999).
\bibitem{LABEL6}
M. Lubo
, Phys.Rev. D {\bf{61}}124009,(2000)
\bibitem{LABEL7}
A.Hebecker, S.Schrechenberg, J.Schwenk, J.Weich, J.Wess, Preprint
MPI-Ph/93-45, LMU-TPW 93-17, Z. Phys. C.
\bibitem{LABEL8}
A. Hebecker, W. Weich, Lett.Math.Phys.{\bf{26}},245(1992).
\bibitem{LABEL9}
T.H. Koornwinder, R.F. Swart touw, Trans.AMS 3311, {\bf{445}d} (1992).
\bibitem{LABEL10}
M. Pillin, W.B. Schmidke, J. Wess, Nucl. Phys. B 403, 223 (1993).
\bibitem{LABEL11}
J. Schwenk, J. Wess, Phys. Lett. B {\bf{291}}, 273 (1992).
\bibitem{LABEL12}
F. Brau, J. Phys. A : Math.Gen.{\bf{32}}, 7691 (1999).
\bibitem{LABEL13}
L.E.Reichl, A modern course in statistical Physics, Edward Arnold, (1980)
\bibitem{LABEL14}
W.Greiner, L.Neise, H.Stocker: Thermodynamics and Statistical Physics,
Springer Verlag, (1995)
\bibitem{LABEL15}
A.Kempf, hep-th/9810215, talk presented at the $ \bf{36^{th}}$ school of subnuclear physics, Reice, Sicily, Sept.98
\bibitem{LABEL16}
N.A.Ahmed, S.Dimopulos, G.Dvali, Phys.Lett.{\bf{B436}},259(1998)
\bibitem{LABEL17}
A.D.Linde , Particle, Particle Physics and Inflationnary cosmology, Harwood Academic Press, Chur, Switzerland (1990)
\bibitem{LABEL18}
M. Pillin, J.Math.Phys.{\bf{35}},2804(1994)
\bibitem{LABEL19}
J. Wess, B. Zumino, Nucl.Phys.{\bf{B18}} (Proc. Suppl.), 302 (1991 ).
\bibitem{LABEL20}
A. Lorek, A. Ruffing, J. Wess, Z.Phys.{\bf{C74}}369(1997)
\bibitem{LABEL21}
  J.Martin, R.H.Brandenberger, hep-th/0005209
\bibitem{LABEL22}
  D.Amati, M.Cialfoni, G.Veneziano , Phys.Lett.{\bf{B216}},41(1989)
\bibitem{LABEL23}
  S.L.Adler , A.Kempf  , J.Math.Phys.{\bf{39}}(1998)5083-5097
\end{thebibliography}
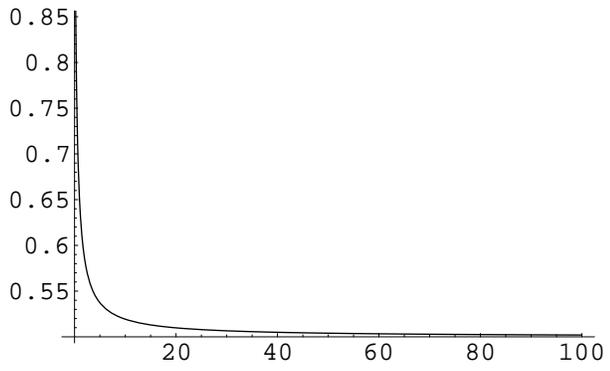
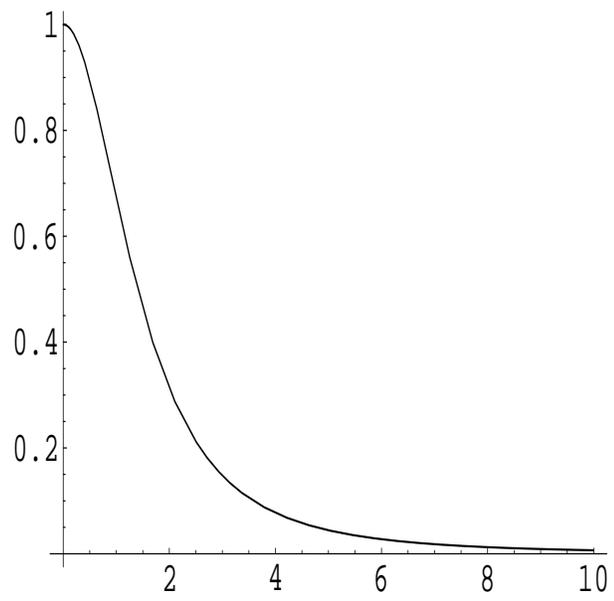
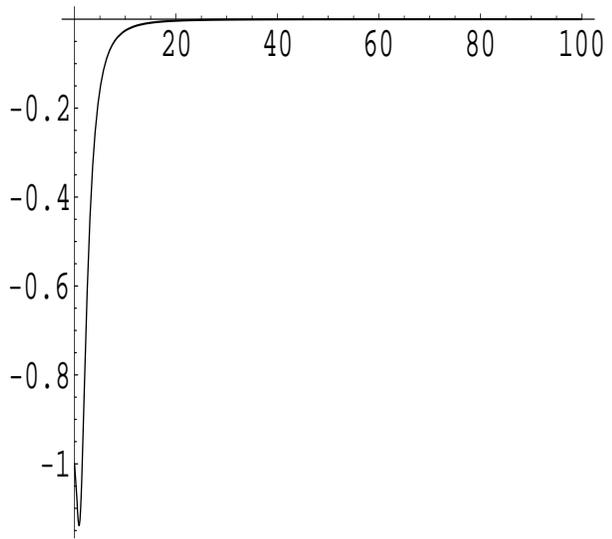
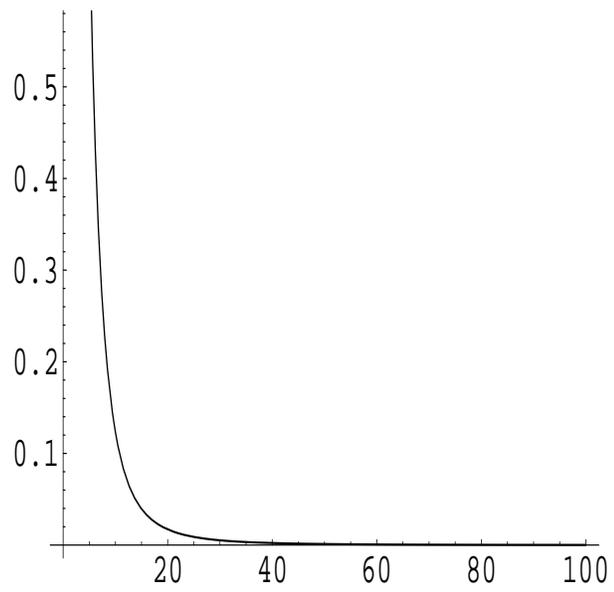
\end{document}